\documentclass[aps,prl,twocolumn,showpacs,groupedaddress]{revtex4}
\usepackage{graphicx}  
\usepackage{dcolumn}   
\usepackage{bm}        
\usepackage{amssymb}   
\usepackage{verbatim}

\usepackage{amsmath}
\usepackage{amsfonts}

\newcommand{\ket}[1]{|#1\rangle}

\begin{document}

\title{Magnetism in SQUIDs at Millikelvin Temperatures}
\author{S. Sendelbach$^1$}
\author{D. Hover$^{1}$}
\author{A. Kittel$^{2}$}
\author{M. M\"{u}ck$^{3}$}
\author{John M. Martinis$^{4}$}
\author{R. McDermott$^{1,}$}
\email[Electronic address: ]{rfmcdermott@wisc.edu}

\affiliation{$^{1}$Department of Physics, University of
Wisconsin-Madison, Madison, Wisconsin 53706, USA}

\affiliation{$^{2}$Institut f\"{u}r Physik, Carl von Ossietzky
Universit\"{a}t, D-26111 Oldenburg, Germany}

\affiliation{$^{3}$Institut f\"{u}r Angewandte Physik, Justus-Leibig-Universit\"{a}t Gie{\ss}en,
D-35392 Gie{\ss}en, Germany}

\affiliation{$^{4}$Department of Physics, University of California,
Santa Barbara, California 93106, USA}

\date{\today}

\begin{abstract}
We have characterized the temperature dependence of the flux threading dc SQUIDs cooled to millikelvin temperatures. The flux increases as $1/T$ as temperature is lowered; moreover, the flux change is proportional to the density of trapped vortices. The data is compatible with the thermal polarization of surface spins in the trapped fields of the vortices. In the absence of trapped flux, we observe evidence of spin-glass freezing at low temperature. These results suggest an explanation for the ``universal" $1/f$ flux noise in SQUIDs and superconducting qubits.
\end{abstract}

\maketitle

Superconducting integrated circuits are a leading candidate for scalable quantum information processing \cite{Devoret}. A key qubit figure of merit is the dephasing time, the time over which a superposition of the qubit $\ket{0}$ and $\ket{1}$ states maintains phase coherence. Dephasing is governed by low-frequency noise in the bias parameters that control the energy separation between the qubit states \cite{Bias,DvH}. In the Josephson flux qubit, the gaussian decay envelope of qubit Ramsey fringes is compatible with a magnetic flux noise with $1/f$ spectrum and a magnitude at 1 Hz of 1 $\mu\Phi_0$/Hz$^{1/2}$, where $\Phi_0=h/2e$ is the magnetic flux quantum \cite{Nakamura,Kakuyanagi}. In the Josephson phase qubit, a recent experiment used the resonant response of the qubit to measure the power spectral density of magnetic flux noise directly; again the spectrum was $1/f$, with a noise magnitude comparable to that seen in the flux qubit \cite{Bialczak}. While these experiments identify flux noise as a dominant dephasing mechanism, they offer no clue to its microscopic origin.

The flux noise inferred from recent qubit experiments is consistent with the flux noise observed more than 20 years ago in a series of measurements performed on dc Superconducting QUantum Interference Devices (SQUIDs) cooled to millikelvin temperatures \cite{Wellstood}. The noise was observed to be ``universal", that is, only weakly dependent on a wide range of parameters such as superconducting materials, SQUID loop geometry and inductance, and temperature. While these experiments ruled out many potential sources of noise, the microscopic mechanism was never identified. There has been recent theoretical interest in the possibility that the $1/f$ flux noise is due to unpaired spins on the surfaces of the superconductors \cite{Koch, de Sousa}. Models for $1/f$ flux noise from surface spins are attractive, as they yield a noise power that is only weakly dependent on the overall scale of the device, compatible with the ``universal" character of the noise \cite{Bialczak}. To date, however, there has been no direct experimental evidence for magnetism in superconducting circuits.
\begin{figure}[b]
\includegraphics[width=.47\textwidth]{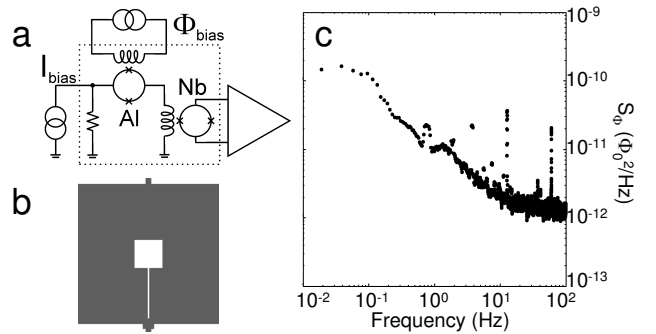}
\vspace*{-0.0in} \caption{(a) Schematic of two-stage SQUID amplifier. The first-stage Al SQUID is biased with a voltage, and the current through this device is read out with a Nb SQUID operated in a flux-locked loop. Both devices are operated inside a superconducting aluminum shield (dashed box). (b) Layout of the thin-film SQUIDs. The Al SQUID has inner and outer dimensions 60 and 300 $\mu$m; the Nb SQUID has inner and outer dimensions 200 $\mu$m and 1 mm. (c) Power spectral density of flux noise of the two-SQUID amplifier measured at 100 mK.}
\label{fig:figure1}\end{figure}

Here we present clear evidence for a high density of unpaired surface spins in thin-film SQUIDs. While fluctuations of these spins produce a small but measurable effect, the coherent magnetization of the spins couples a very large flux of order 1 $\Phi_0$ to the SQUID. Moreover, simple measurements of the temperature dependence of the magnetization allow determination of the surface density of unpaired spins, and indicate that interactions between spins are significant.

We first characterized the flux noise of a two-stage SQUID amplifier at millikelvin temperatures. A schematic of the experiment is shown in Fig. 1. Both SQUIDs are made in the conventional Ketchen-Jaycox square-washer geometry (see inset) \cite{Ketchen}. 
The 110 pH Al SQUID is biased with a voltage, and the current through the SQUID is monitored with the 350 pH Nb SQUID, which is operated in a flux-locked loop \cite{SQUID}. The current through the first-stage Al SQUID oscillates with applied flux, with periodicity $\Phi_0$. When the first-stage SQUID is biased at a flux $(n\pm1/4)\Phi_0$, where $n$ is an integer, sensitivity to applied flux is maximum.

The power spectral density of flux noise in the Al first-stage SQUID is shown in Fig. 1. The noise scales as $1/f$ at low frequencies, with a noise magnitude of 3 $\mu \Phi_0$/Hz$^{1/2}$ at 1 Hz. The flux noise is roughly independent of temperature over the range from 30 mK to 500 mK (not shown). These results are compatible with previous measurements of SQUIDs and qubits \cite{Wellstood,Nakamura,Kakuyanagi,Bialczak}.

At the lowest temperatures, however, we observe a surprising dependence of SQUID flux on bath temperature. We characterize the temperature dependence of the flux in the SQUIDs by tracing out the full current-flux ($I-\Phi$) characteristic of the first-stage SQUID as bath temperature is varied. By following shifts in the position of the extrema of the $I-\Phi$ curve along both the $\Phi$ and $I$ axes, we monitor changes in the quasistatic fluxes threading the first- and second-stage SQUIDs, respectively. The results of this experiment are shown in Fig. 2.
\begin{figure}[t]
\includegraphics[width=.47\textwidth]{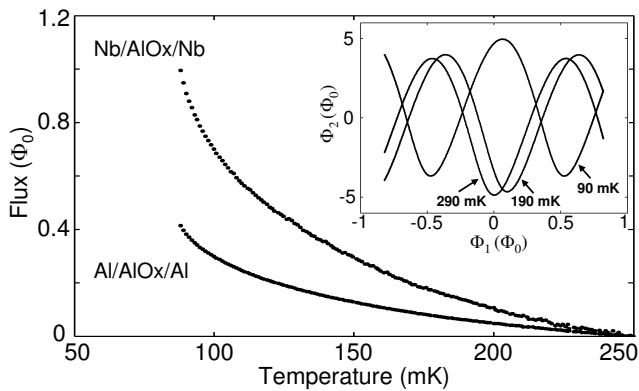}
\caption{Temperature dependence of flux threading the SQUIDs of the amplifier described in Fig. 1. Inset shows the $I-\Phi$ characteristic of the Al first-stage SQUID measured at three temperatures; flux $\Phi_1$ through the Al first-stage SQUID is plotted on the horizontal axis, and current $I$ through the Al first-stage SQUID is expressed as a flux $\Phi_2$ coupled to the Nb second-stage SQUID and plotted on the vertical axis.}
\label{fig:figure2}
\end{figure}

The fluxes threading the SQUIDs change dramatically as bath temperature is lowered. At the lowest accessible temperatures $\sim$ 90 mK, the temperature-to-flux transfer coefficients are in excess of 10 $\Phi_0$/K. This behavior has been observed in both Al and Nb devices, made both with and without a wiring dielectric, and prepared in different facilities according to different fabrication recipes.

Temperature-induced flux drift in SQUIDs has been studied in the past. There have been investigations of flux shifts due to thermally-driven motion of magnetic flux \cite{Clarke, Mueck}, and due to temperature-dependent critical currents in asymmetric SQUIDs \cite{Clarke}. However, we do not expect these mechanisms to play a role here, since our experiments are conducted far below the superconducting transition temperature $T_c$. Indeed, the large magnitude and low energy scale of the effect observed here suggest a new mechanism. Over a broad temperature range below 500 mK, the flux threading the SQUIDs scales inversely with temperature.  Paramagnetic impurities in the materials of the SQUID would naturally give rise to such a signature, and we interpret the $1/T$ dependence of the flux as strong evidence for unpaired spins, most likely in the native oxides of the superconductors. However, as these experiments are performed in a nominal zero field, the source of the orienting magnetic field is not immediately clear.

In order to clarify the source of the temperature-dependent flux, we have performed a series of field-cool experiments in which a magnetic field $B_{fc}$ is applied to a 350 pH square-washer Nb SQUID as it is cooled through $T_c$; the field freezes magnetic flux vortices into the Nb film, with density  $\sigma_v\approx B_{fc}/\Phi_0$ \cite{Stan}. When the device is well below $T_c$, the field is removed, and the SQUID is maintained in a flux-locked loop as it is cooled to millikelvin temperatures. In Fig. 3a we plot the flux through the SQUID as a function of temperature for eight different values of the cooling field. The cooling field strongly affects the temperature-dependent flux, enhancing or even reversing the polarity of the observed signal. In Fig. 3b we plot the flux change on cooling from 500 mK to 100 mK as a function of the cooling field; a linear fit to the data yields a slope of 1.3 $\Phi_0$/mT. Clearly, vortices contribute significantly to the measured temperature-induced flux shift.

\begin{figure}[b]
\includegraphics[width=.47\textwidth]{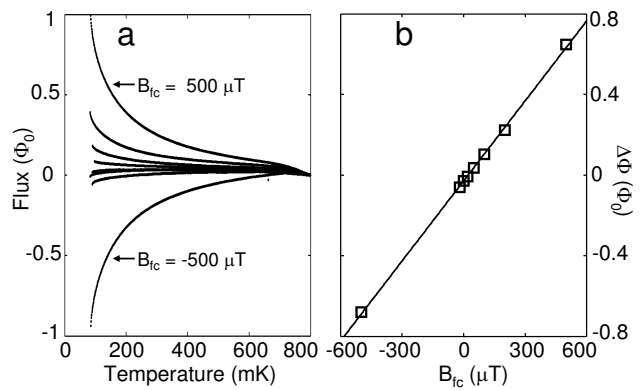}
\vspace*{-0.0in} \caption{(a) Temperature dependence of the flux threading a 350 pH Nb/AlOx/Nb SQUID, for different fields $B_{fc}$ applied as the device was cooled through the superconducting transition (from bottom to top, $B_{fc}$ = -500, -20, 0, 20, 50, 100, 200, 500 $\mu$T). (b) Temperature-induced flux change $\Delta \Phi = \Phi(100 \textrm{mK}) - \Phi(500 \textrm{mK})$ on cooling from 500 mK to 100 mK \textit{vs.} cooling field $B_{fc}$. A linear fit to the data yields a slope $\Delta \Phi/B_{fc}$ = 1.3 $\Phi_0$/mT.}
\label{fig:figure3}\end{figure}

The linear dependence of flux drift on vortex density suggests the following interpretation of the data. As the superconducting films are cooled through $T_c$ in an applied field, vortices nucleate and relatively large magnetic fields of order 10 mT are frozen into the films. The observed signal is due to the magnetization of unpaired electron spins on the surface of the superconductor in the strong fields produced by the trapped vortices. Indeed, careful analysis of the field-cool data allows us to extract the surface density $\sigma_s$ of spins, a key parameter in models of $1/f$ noise from surface magnetism \cite{McDermott}. We note that there is negligible \textit{direct} coupling to the SQUID from spins polarized out of the plane of the superconducting films; this is easily understood from reciprocity, as surface magnetic fields due to currents in the SQUID have vanishing perpendicular component. However, the topology of our device is quite different from that of a continuous superconducting washer, due to the presence of the vortices. The thermal polarization of unpaired surface spins in the vortex forces a redistribution of the circulating supercurrents in the vortex, due to the requirement to conserve magnetic flux. These currents, in turn, couple strongly to the SQUID loop. Calculation of the spin density from the data of Fig. 3 therefore proceeds in two stages: (1) calculation of the coupling between a vortex and the SQUID loop, and (2) calculation of the coupling between surface spins and a vortex. (A detailed discussion of this calculation can be found in \cite{McDermott}). 

A single vortex trapped in the SQUID washer induces an effective flux offset in the SQUID; the magnitude of the flux offset approaches 0 $\Phi_0$ and 1 $\Phi_0$ for vortices positioned at the outer and inner edges of the SQUID washer, respectively.  Numerical calculations for a circular washer show that this effective flux offset averaged over the area of the washer is 0.5 $\Phi_0$ for a thin washer (with width much less than radius), as expected; for our device dimensions (inner radius 100 $\mu$m and outer radius 500 $\mu$m), we find an average flux offset of 0.14 $\Phi_0$. The physical source of the flux is vortex currents that drop off slowly with distance from the core and circulate around the hole in the SQUID washer. Polarization of the surface spins couples a flux $\Phi_v(T)$ to the vortex, requiring the circulating currents to change due to the quantization condition.  The change in flux coupled to the SQUID is thus $\Phi = -0.14 \Phi_v(T)$.

To calculate $\Phi_v(T)$, we rely on reciprocity: knowledge of the current distribution in the vortex allows us to determine the coupling to a spin at an arbitrary location. We model the vortex as a hole in a superconducting groundplane with radius equal to the estimated coherence length $\xi$ = 30 nm and a thin-film penetration depth $\Lambda$ = 100 nm. Numerical solution of the London equations allows us to solve for the currents in the vortex, from which we determine the in-plane and out-of-plane magnetic fields $B_r(r)$ and $B_z(r)$. One can show that the flux coupled to the vortex by the spins is given by
\begin{align}
\Phi_v(T) = \mu_B \sigma_s L_v P_{eff}(T),
\label{eq:spinvortex}
\end{align}
where $L_v$ is the vortex self-inductance, and where we have defined an effective spin polarization $P_{eff}(T)$:
\begin{align}
P_{eff}(T) = \frac{1}{\Phi_0} \displaystyle \int^\infty_0 \,2\pi r B(r) \tanh \left(\frac{\mu_B B(r)}{2k_BT}\right)\,dr,
\label{eq:spinvortex}
\end{align}
where $B(r) = \left[B_r(r)^2 + B_z(r)^2\right]^{1/2}.$ A change in temperature yields a change in effective spin polarization $P_{eff}(T)$ in the vortex, which in turn induces a change in flux $\Phi_v(T)$ coupled to each vortex. The total flux change $\Delta \Phi$ at the SQUID is obtained by summing over all vortices, taking into account the coupling factor 0.14. Thus, we can relate the measured temperature-dependent flux to spin density as follows:
\begin{align}
\frac{\Delta \Phi}{B_{fc}} = 0.14 \frac{A_{SQ}}{\Phi_0} \, \mu_B \sigma_s L_v \Delta P_{eff},
\label{eq:totalFlux}
\end{align}
where $A_{SQ}$ is the area of the SQUID washer. For the materials parameters given above, we find $\Delta P_{eff} \equiv P_{eff}\left(100 \, \textrm{mK}\right) - P_{eff}\left(500 \, \textrm{mK}\right) = 0.037$ and $L_v$ = 0.24 pH. Using $A_{SQ}$ = 0.96 mm${^2}$ and the slope $\Delta \Phi/B_{fc}$ = 1.3 $\Phi_0$/mT from Fig. 3b, we find a spin density $\sigma_s$ = $5.0 \times 10^{17}$ m$^{-2}$. This density of surface spins is compatible with densities considered in recent theoretical models of $1/f$ flux noise from surface spins \cite{Koch,de Sousa}, and with the surface spin density postulated in a recent model for spin-flip scattering that has been proposed \cite{Wei} to account for the magnetic-field enhancement of the critical current of superconducting nanowires \cite{Rogachev}.

Yet while spin densities of this order of magnitude are clearly required to explain the measured $1/f$ flux noise and qubit dephasing times in terms of surface magnetism, the microscopic physics that drives the spin fluctuations is by no means obvious. A recent theory explains flux noise in terms of spin diffusion and interaction between surface spins via the RKKY mechanism \cite{Faoro}. A key parameter in this model is the characteristic energy of spin interactions. In the field-cool experiments described above, any possible spin interactions are obscured by strong coupling of the spins to the fields of the vortices. To probe interactions between surface spins, it is necessary to examine devices that have no trapped flux. To this end, we have performed additional experiments using SQUIDs made from narrow (2 $\mu$m) superconducting lines. It is energetically unfavorable for a vortex to nucleate in a superconducting strip of width $w$ provided the strip is cooled through $T_c$ in a magnetic field that is less than the threshold field $B_{thr} \approx \Phi_0/w^2$ \cite{Clem, Dantsker, Stan}. As the threshold field for our narrow-linewidth SQUIDs is around 500 $\mu$T, we expect no trapped vortices when these devices are cooled in a nominal zero field.
\begin{figure}[t]
\includegraphics[width=.47\textwidth]{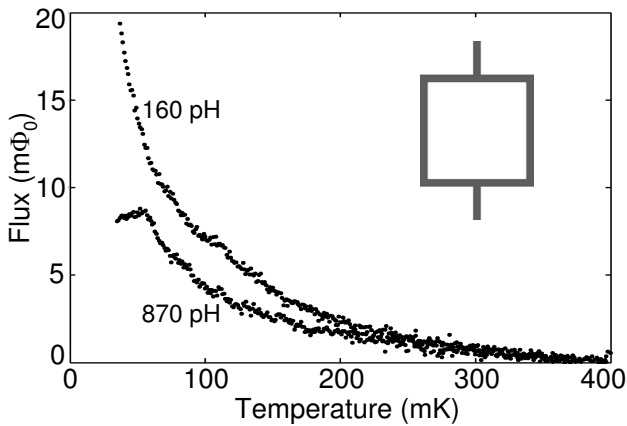}
\vspace*{-0.0in} \caption{Temperature dependence of the flux threading narrow (2 $\mu$m) linewidth Nb/AlOx/Nb SQUIDs with inductances 160 pH and 870 pH. Device geometry is shown in the inset.}
\label{fig:figure4}\end{figure}

In Fig. 4 we show typical flux \textit{vs.} temperature data for two narrow linewidth Nb/AlOx/Nb SQUIDs configured as square loops with dimensions $d$ = 50 $\mu$m and 200 $\mu$m (see inset). The narrow devices show a strongly temperature-dependent flux even in the absence of trapped vortices. Although the exact magnitude of the total flux change is seen to vary from cooldown to cooldown, suggesting a dependence of the magnetization on the details of the cooldown, 
we consistently observe a flux shift of order 10 m$\Phi_0$ on cooling to 30 mK. We take this as confirmation that the flux drift observed in our experiments is not due to thermally active vortices. Moreover, these measurements allow us to extract an energy scale of spin interactions, as we explain below.

First, the magnitude of the temperature-dependent flux in these devices indicates a high degree of spin order. Following the analysis of \cite{Bialczak}, one can show that the maximum flux coupled to the SQUID from surface spins (achieved when the spins are fully polarized and oriented radially) is given by
\begin{align}
\Phi_{max} = 2\mu_0 \mu_B \sigma_s d.
\end{align}
The flux scales linearly with device dimension, and is independent of the width of the superconducting lines for $w<<d$. For the 160 pH SQUID with $d$ = 50 $\mu$m, a spin density $\sigma_s = 5 \times 10^{17}$ m$^{-2}$ yields $\Phi_{max}$ = 280 m$\Phi_0$; the measured flux change of 19 m$\Phi_0$ therefore suggests a high degree of spin polarization at 30 mK. While surface spin densities in the narrow SQUIDs could be higher than that inferred from the field-cool experiments (the narrow devices are not passivated by a wiring dielectric), the large temperature-induced flux shift in these devices in the absence of a strong applied field suggests an energy scale for spin interactions of order 10 mK. Pure dipolar coupling at the spin densities considered here would yield a much lower energy scale, of order 100 $\mu$K. On the other hand, the RKKY model of \cite{Faoro} predicts a spin interaction strength of 20 mK for the spin densities and materials parameters considered here, assuming a uniform distribution of spins through the native oxide; for spins concentrated at the metal-insulator interface, the spin interaction strength approaches 500 mK.

Moreover, we note the appearance of a cusp at 55 mK in the flux \textit{vs.} temperature data of the 870 pH SQUID. This is reminiscent of the cusp in the first-order susceptibility of a spin glass at the freezing temperature \cite{Binder}. This feature again indicates the importance of spin interactions at millikelvin temperatures and suggests the possibility of spin-glass freezing of the unpaired surface spins on the timescales of our experiment.

In summary, we have demonstrated that superconducting circuits are magnetically active at millikelvin temperatures. This observation points to a microscopic explanation for the excess $1/f$ flux noise in Josephson circuits, a dominant source of dephasing in superconducting qubits and an open question for more than 20 years. We observe evidence of interactions between surface spins and possible spin-glass freezing at millikelvin temperatures. A high density of surface spins could play a deleterious role in other schemes for quantum information processing in the solid-state, notably those based on electron spin. It is possible that suitable surface treatments of the superconducting films will lower the density of magnetic states, leading to superconducting devices with lower noise and solid-state qubits with improved coherence times.

\begin{acknowledgments}
We acknowledge useful discussions with L. Faoro, L.B. Ioffe, B.L.T. Plourde, and C.C. Yu. Some devices were fabricated at the UCSB Nanofabrication Facility, part of the NSF-funded NNIN. This work was supported in part by the U.S. Government. The views and conclusions contained in this document are those of the authors and should not be interpreted as representing the official policies, either expressly or implied, of the U.S. Government.
\end{acknowledgments}

\end{document}